\documentclass[journal,twoside,web,letterpaper]{ieeecolor}
\usepackage{lcsys}

\usepackage{savesym}
\usepackage{xcolor}
\usepackage{amsmath,amssymb,amsfonts}
\savesymbol{AND}
\usepackage{algorithm}
\usepackage{algorithmic}
\usepackage{setspace}
\usepackage{caption}
\usepackage{subfigure}
\usepackage{verbatim}
\usepackage{enumerate}
\usepackage{url}

\usepackage{tikz}
\usepackage{multirow}
\usepackage{booktabs}

\usepackage{bm} 
\newtheorem{theorem}{Theorem}
\newtheorem{lemma}{Lemma}
\newtheorem{problem}{Problem}
\newtheorem{example}{Example}
\newtheorem{definition}{Definition}

\title{\LARGE \bf
Transformational Supervisor Localization
}

\author{Sander Thuijsman$^a$, Kai Cai$^b$, and Michel Reniers$^a$
\thanks{$^{a}$Eindhoven University of Technology, Eindhoven, Netherlands; {\tt $\lbrace$s.b.thuijsman, m.a.reniers$\rbrace$@tue.nl}}%
\thanks{$^{b}$Osaka Metropolitan University, Osaka, Japan; {\tt cai@omu.ac.jp}}%
\thanks{This work was supported in part by JSPS KAKENHI Grant nos. 21H04875, 22KK0155}
}

\usepackage{eso-pic}

\begin{document}

\maketitle
\thispagestyle{empty}
\pagestyle{empty}

\AddToShipoutPictureBG*{%
\AtPageUpperLeft{%
\setlength\unitlength{1in}%
\hspace*{\dimexpr0.5\paperwidth\relax}
\makebox(0,-1)[c]{
\begin{tabular}{c c}
Sander Thuijsman \emph{et al.},
Transformational Supervisor Localization, \\
Accepted for: {\em IEEE Control Systems Letters (L-CSS)} (2023)\\
Uploaded to ArXiv \today. \\
\end{tabular}}}}

\AddToShipoutPictureBG*{%
\AtPageUpperLeft{%
\setlength\unitlength{1in}%
\hspace*{\dimexpr0.5\paperwidth\relax}
\makebox(0,-21.5)[c]{
\footnotesize
\begin{tabular}{c c}
© 2023 IEEE. Personal use of this material is permitted. Permission from
IEEE must be obtained for all other uses, in any \\
current or future media, including reprinting/republishing
this material for advertising or promotional purposes, creating new \\
collective works, for resale or redistribution to servers or lists,
or reuse of any copyrighted component of this work in other works.
\end{tabular}}}}

\begin{abstract}
Supervisor localization can be applied to distribute a monolithic supervisor into local supervisors.
Performing supervisor localization can be computationally costly.
In this work, we consider systems that evolve over time.
We study how to reuse the results from a previous supervisor localization, to more efficiently compute local supervisors when the system is adapted.
We call this approach transformational supervisor localization, and present algorithms for the procedure.
The efficiency of the procedure is experimentally evaluated.
\end{abstract}
\begin{IEEEkeywords}
discrete-event systems, supervisory control, model/controller reduction, computational methods, automata
\end{IEEEkeywords}


\section{Introduction}
\IEEEPARstart{S}{upervisory} control theory, as introduced by Ramadge and Wonham \cite{Ramadge1987}, is a model-based approach to control discrete event (dynamic) systems.
Typically, cyber-physical systems are modeled.
By applying \textit{supervisor synthesis} on a model of an uncontrolled system (plant) and system requirements, a correct-by-construction supervisor is obtained.
This supervisor enables/disables events such that the requirements are always adhered to, and some more behavioral properties apply to the controlled system such as: nonblockingness, controllability, and maximal permissiveness \cite{Cassandras2021}.
The most straightforward approach is \textit{monolithic} supervisor synthesis, which computes a single global supervisor that controls all components and enforces all requirements. 

Large, global controllers may be undesirable in practice.
As such, many modern control systems are distributed over a number of agents \cite{Moormann2021}. 
These agents may act locally based on their own observations and control strategies.
Through \textit{supervisor localization} (SL) \cite{Cai2010}, local supervisors for the individual agents are computed from the monolithic supervisor, that together achieve the same controlled behavior as the monolithic supervisor.
SL is an extension to \textit{supervisor reduction}, which converts a supervisor automaton to a smaller automaton (with less states) that is control equivalent to the original automaton \cite{Su2004}.
We present preliminaries on supervisor localization/reduction in Section \ref{sec:preliminaries}.

In this work, we slightly modify the SL algorithm from \cite{Su2004,Cai2010} to be able to initialize it in a way such that it has to do less calculations/loops, which benefits the method that we are going to introduce.
Furthermore, because it is desirable to obtain small (in terms of number of states) local supervisors, we show that the local supervisors obtained by SL are maximally reduced.
These novel extensions to SL are presented in Section \ref{sec:SL}.

In recent work, transformational approaches for supervisory control algorithms, such as transformational supervisor synthesis \cite{Thuijsman2022} and transformational nonblocking verification \cite{Thuijsman2022b}, are investigated. 
Such transformational approaches deal with cyber-physical systems that evolve over time.
Results of previous computations, such as synthesis or verification, may not be valid anymore once the system is adapted.
In this case, tranformational methods can be applied that reuse the output of previous calculations to more efficiently compute the result of some algorithm, rather than computing it from scratch.
The general idea is that the previous result is \textit{transformed} into the new result, using knowledge on how the system is adapted.


In this paper, we investigate \textit{transformational supervisor localization} (TSL).
We assume a \textit{base model}, on which (T)SL has already been performed.
The base model is adapted such that a \textit{variant model} is obtained.
The goal is to use the localization output of the base model, to more efficiently compute local supervisors for the variant model.
The formal problem definition is given in Section \ref{sec:problem}.
We present algorithms for TSL and prove their correctness in Section \ref{sec:TSL}.
The computational benefit of TSL is evaluated by a use case in Section \ref{sec:CMT}.


\section{Preliminaries}
\label{sec:preliminaries}
In the following we discuss the preliminaries on SL.
We first provide automata definitions for plant and (monolithic) supervisor.
The plant is assumed to be a composition of agents.
The goal is to generate local supervisors that each supervise an agent.
This is done by grouping states of the monolithic supervisor into cells, that are consistent in the enablement and disablement of events controlled by the respective agent.
These cells are the states of the local supervisor.
Together, the behavior of the system under control by the local supervisors is the same as that of the monolithic supervisor.
For details we refer to \cite{Cai2010}.

The plant is defined by a finite state automaton $G=(Q,\Sigma,\delta,q_0,Q_m)$, where $Q$ is the finite state set, $\Sigma$ is the finite event set, $q_0$ is the initial state, and $Q_m$ is the subset of marked states. 
$\delta:Q\times\Sigma\rightarrow Q$ is the (partial) transition function.
We denote $\delta(q,\sigma)!$ if $\delta(q,\sigma)$ is defined.
We extend this notation to $\delta:Q\times\Sigma^*\rightarrow Q$, and write $\delta(q,s)$ for $s\in \Sigma^*$ to indicate sequences of transitions.
We consider the case that $G$ is formed by a composition of $n$ agents, that each have local events: $\dot\bigcup_{k\in \{1,...,n\}}\Sigma_k=\Sigma$, from which a subset are locally controllable $\Sigma_{c,k}\subseteq \Sigma_k$. 
We assume a monolithic supervisor is provided for plant $G$, defined by finite state automaton $S=(X,\Sigma,\xi,x_0,X_m)$.
For the purpose of the algorithms in this work, we assume the states are numbered/indexed, i.e., $X=\{x_0,x_1,...\}$.


We use the following functions \cite{Cai2010}:
\begin{itemize}
\item $E:X\rightarrow 2^\Sigma$, where $E(x) = \{\sigma\in\Sigma | \xi(x,\sigma)!\}$
\item $D_k: X\rightarrow 2^{\Sigma_{c,k}}$, and $D_k(x) = \{\sigma\in \Sigma_{c,k} | \neg \xi (x,\sigma)! \wedge (\exists s\in \Sigma^*):(\xi(x_0,s)=x\wedge \delta(q_0,s\sigma)!)\}$
\item $M:X\rightarrow \{0,1\}$, where $M(x)=1$ iff $x\in X_m$
\item $T:X\rightarrow \{0,1\}$, where $T(x)=1$ iff $(\exists s\in \Sigma^*):(\xi(x_0,s)=x\wedge \delta(q_0,s)\in Q_m)$
\end{itemize}
$E$ indicates events enabled by the supervisor in state $x$.
$D_k$ indicates the events from $\Sigma_{c,k}$ disabled by the supervisor in state $x$. 
$M$ determines if a state is marked in $S$, and $T$ determines if some corresponding state is marked in $G$.

We define control consistency relation $\mathcal{R}_k\subseteq X\times X$ (for agent $k$): for every $x,x'\in X$, $(x,x')\in \mathcal{R}_k$ iff:
\begin{align}
E(x)\cap D_k(x') =\emptyset= E(x') \cap D_k(x)\\
T(x)=T(x')\implies M(x)=M(x')
\end{align}


Cover $\mathcal{C}_k=\{X_i\subseteq X | i\in I_k\}$ with suitable index set $I_k$ is called a \textit{control cover} with respect to some $\Sigma_k$ iff:
\begin{align*}
(i)&\ (\forall i\in I_k,\forall x,x'\in X_i)(x,x')\in \mathcal{R}_k\\
(ii)&\ (\forall i\in I_k,\forall \sigma\in \Sigma) \big{[} \big{(} (\exists x \in X_i) \xi(x,\sigma)! \big{)} \implies \\
&\ \ \big{(} (\exists j\in I_k)(\forall x'\in X_i)\xi(x',\sigma)!\implies \xi(x',\sigma)\in X_j \big{)} \big{]}
\end{align*}

If a control cover $\mathcal{C}$ is a partition on $X$, it is called a \textit{control congruence}.

In this work we frequently address a \textit{singleton cover} $\mathcal{C}=\{\{x\}|x\in X\}$, which trivially always is a control congruence.

We call a set of states in a cover a \textit{cell}. 
In our notation we use $[x]_{\mathcal{C}}$ to refer to the set of states contained in the same cell as $x$ in cover $\mathcal{C}$, or simply $[x]$ if there is no ambiguity.

Given a control congruence $\mathcal{C}_k$, a local supervisor $LOC_k$ is computed as follows (simplified from \cite{Su2004}): $LOC_k=(\mathcal{C}_k,\Sigma,\eta_k,y_{0,k},Y_{m,k})$, where:
$\eta_k:\mathcal{C}_k\times\Sigma\rightarrow\mathcal{C}_k$, with $\eta_k(\pi_1,\sigma)=\pi_2$ iff $(\exists x\in \pi_1):\xi(x,\sigma)\in\pi_2$; $y_{0,k}=[x_0]$; and $Y_{m,k}=\{[x]|x\in X_m\}$.
A local supervisor is deterministic as a result of condition (\textit{ii}) for the control cover.

The set of local supervisors $\{LOC_k|1{\leq} k{\leq} n\}$ constructed in this way is \textit{control equivalent} to $S$ with respect to $G$ \cite{Cai2010}:
\begin{align}
L(G)\cap \bigcap {}_{1\leq k\leq n}L(LOC_k) &= L(S) \cap L(G)\label{eq:controlequiv1}\\
L_m(G)\cap \bigcap {}_{1\leq k\leq n}L_m(LOC_k) &= L_m(S) \cap L_m(G)\label{eq:controlequiv2}
\end{align}
$L(A)$ and $L_m(A)$ respectively denote the language and the marked language of automaton $A$ \cite{Cassandras2021}.

%
%

\section{Supervisor localization}
\label{sec:SL}
In the process of SL, for each agent, a control congruence is computed and subsequently the local supervisor is generated.
We can use the definitions and functions from Section \ref{sec:preliminaries} to perform the localization algorithm, shown in Algorithm \ref{alg:localization}, which makes calls to Algorithm \ref{alg:checkmerge} \cite{Cai2010}\footnote{Relative to \cite{Su2004,Cai2010} some minor changes have been made to lines 1,2, and 7 of Algorithm \ref{alg:checkmerge} for correctness.}.
Note that, e.g., the $X$ on line 1 implicitly originates from automaton $S$.
A `continue' ends current execution and the function goes to the next iteration of the nearest enclosing for-loop.
A `return' ends current call to the algorithm and the specified values are returned to the parent routine.


\begin{algorithm}[t]
\small
 \caption{\texttt{localize}}
 \label{alg:localization}
 \begin{algorithmic}[1]
 \renewcommand{\algorithmicrequire}{\textbf{Input:}}
 \renewcommand{\algorithmicensure}{\textbf{Output:}}
\renewcommand{\algorithmicendif}{\textbf{end}}
\renewcommand{\algorithmicendfor}{\textbf{end}}
\renewcommand{\algorithmicendwhile}{\textbf{end}}
 \REQUIRE $G$, $S$, $\Sigma_{c,k}$, initial $\mathcal{C}_k$
 \ENSURE Control congruence $\mathcal{C}_k$
 \STATE \textbf{for} $i=0$ to $|X|-2$ \textbf{do}
 \STATE \hspace{1em}\textbf{if} $i>\mathit{min}(\{m|x_m\in [x_i]\})$ \textbf{then continue; end}
 \STATE \hspace{1em}\textbf{for} $j=i+1$ to $|X|-1$ \textbf{do}
 \STATE \hspace{2em}\textbf{if} $j>\mathit{min}(\{m|x_m\in [x_j]\})$ \textbf{then continue; end}
 \STATE \hspace{2em}$W=\emptyset$
 \STATE \hspace{2em}$(\mathit{flag},W)=\text{\texttt{check\_merge}}(x_i,x_j,W,i,\xi,\mathcal{C}_k)$
 \STATE \hspace{2em}\textbf{if} \textit{flag} \textbf{then}
 \STATE \hspace{3em}$\mathcal{C}_k=\Big{\lbrace} [x] \cup \bigcup \{[x'] | \{(x,x'),(x',x)\}\cap W\neq \emptyset \}$
 
 \hspace{19em}$\Big{\vert} [x],[x']\in\mathcal{C}_k \Big{\rbrace}$
 \STATE \hspace{2em}\textbf{end}
 \STATE \hspace{1em}\textbf{end}
 \STATE \hspace{0em}\textbf{end}
 \STATE \textbf{return} $\mathcal{C}_k$
 \end{algorithmic}
\end{algorithm}

\begin{algorithm}[t]
\small
 \caption{\texttt{check\_merge}}
 \label{alg:checkmerge}
 \begin{algorithmic}[1]
 \renewcommand{\algorithmicrequire}{\textbf{Input:}}
 \renewcommand{\algorithmicensure}{\textbf{Output:}}
\renewcommand{\algorithmicendif}{\textbf{end}}
\renewcommand{\algorithmicendfor}{\textbf{end}}
\renewcommand{\algorithmicendwhile}{\textbf{end}}
 \REQUIRE $x_i$, $x_j$, waiting list $W$, $i$, $\xi$, $\mathcal{C}_k$
 \ENSURE mergeability Boolean $\mathit{flag}$, $W$
 \STATE \textbf{for all} $x_p\in [x_i]\cup\bigcup \{[x]|\{(x,x_i'),(x_i',x)\}\cap W\neq\emptyset,$
 
 \hspace{19.6em}$x_i'\in [x_i]\}$ \textbf{do}
 \STATE \hspace{1em}\textbf{for all} $x_q\in [x_j]\cup\bigcup \{[x]|\{(x,x_j'),(x_j',x)\}\cap W\neq\emptyset,$
 
 \hspace{19.5em}$x_j'\in [x_j]\}$ \textbf{do}
 \STATE \hspace{2em}\textbf{if} $\{(x_p,x_q),(x_q,x_p)\}\cap W\neq \emptyset$ \textbf{then continue; end}
 \STATE \hspace{2em}\textbf{if} $(x_p,x_q)\not\in \mathcal{R}_k$ \textbf{then return} $(\mathit{false},W)$\textbf{; end}
 \STATE \hspace{2em}$W=W\cup \{(x_p,x_q)\}$
 \STATE \hspace{2em}\textbf{for all} $\sigma\in E(x_p)\cap E(x_q)$
 \STATE \hspace{3em}\textbf{if} $[\xi(x_p,\sigma)]=[\xi(x_q,\sigma)]$ \textbf{or} 
 
 \hspace{2.5em} $\{(\xi(x_p,\sigma),\xi(x_q,\sigma)),(\xi(x_q,\sigma),\xi(x_p,\sigma))\}\cap W\neq \emptyset$ 
 
 \hspace{16.5em} \textbf{then continue; end}
 \STATE \hspace{3em}\textbf{if} \textit{min}$(\{m|x_m\in [\xi(x_p,\sigma)]\})<i$ \textbf{or}
 
 \hspace{12em}\textit{min}$(\{m|x_m\in [\xi(x_q,\sigma)]\})<i$ 
 
 \hspace{13em}\textbf{then return} $(\mathit{false},W)$\textbf{; end}
 \STATE \hspace{3em}$(\mathit{flag},W)=\text{\texttt{check\_merge}}(\xi(x_p,\sigma),\xi(x_q,\sigma),W,i,$
 
 \hspace{23.2em}$\xi,\mathcal{C}_k)$
 \STATE \hspace{3em}\textbf{if not} $\mathit{flag}$ \textbf{then return} $(\mathit{false},W)$\textbf{; end}
 \STATE \hspace{2em}\textbf{end}
 \STATE \hspace{1em}\textbf{end}
 \STATE \hspace{0em}\textbf{end}
 \STATE \textbf{return} $(\mathit{true},W)$
 \end{algorithmic}
\end{algorithm}

\begin{example}
\label{ex:ex1}
We consider the supervisor automaton shown in Fig. \ref{Ex1S}.
The states are represented by circles. 
The dangling incoming arrow indicates $x_0$ is the initial state.
Transitions are shown by arrows between states with the respective event label.
To simplify the examples, no states are marked and all events are controllable.

\begin{figure}[t]
\centering
\subfigure[Supervisor]{
\includegraphics[scale=1]{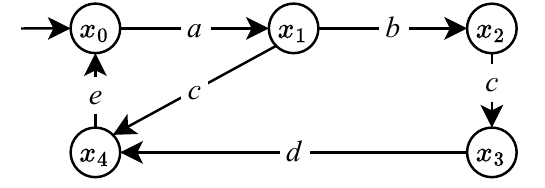}
\label{Ex1S}
}
\subfigure[Local supervisor]{
\includegraphics[scale=1]{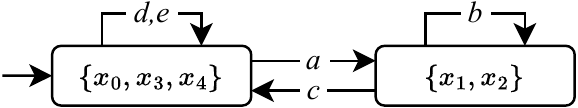}
\label{Ex1LOC}
}
\caption{Automata of Example 1}
\label{Ex1}
\end{figure}

We consider the case that there is an agent (numbered 1) whose set of local controllable events includes all events, i.e., $\Sigma_{c,1}{=}\Sigma_{1}{=}\Sigma{=}\{a,b,c,d,e\}$.
Let us consider the case that the supervisor disables event $c$ in state $x_0$, and disables event $a$ in state $x_2$.
There are no disablements in the other states, i.e., the supervisor permits the same events as the plant in those states.
So, $D_1(x_0){=}\{c\}, D_1(x_2){=}\{a\}$, $D_1(x_1){=}D_1(x_3){=}D_1(x_4){=}\emptyset$.

To compute the local supervisor, we perform the localization algorithm initialized with a singleton cover $\{\{x_0\},...,\{x_4\}\}$.
First, mergeability of $x_0$ and $x_1$ is checked. 
These states are not mergeable, since event $c$ is disabled in $x_0$ but enabled in $x_1$.
Also $x_0$ and $x_2$ are not mergeable.
$x_0$ is mergeable with $x_3$ and they are subsequently merged.
Next, $\{x_0,x_3\}$ is merged with $x_4$ to form cell $\{x_0,x_3,x_4\}$.
Finally $x_1$ and $x_2$ are merged, and no more merges are possible so the algorithm terminates.
Using the resulting control congruence, a local supervisor is constructed, which is displayed in Fig. \ref{Ex1LOC}.
\null\nobreak\hfill\ensuremath{\blacklozenge}
\end{example}

In \cite{Cai2010} the localization algorithm is initiated with a singleton cover.
However, in this work we will also initialize the algorithm with non-singleton covers, to benefit the efficiency of the transformational method that we are going to introduce.
We present Lemma \ref{lm:initialization} on this initialization.

\begin{lemma}
\label{lm:initialization}
If Algorithm \ref{alg:localization} is initiated with a control congruence $\mathcal{C}_{k,init}$, the output cover $\mathcal{C}_k$ is a control congruence.

\begin{proof}
Correctness of Algorithm \ref{alg:localization} initiated by a singleton cover is proven in \cite{Su2004}. 
The singleton cover is a special instance of a control congruence.
The same proof of \cite{Su2004} applies here, when we generalize the algorithm to be initialized with any control congruence.
\end{proof}
\end{lemma}

It is desirable to have small (in terms of number of states) local supervisors. 
Therefore, we want to compute control congruences which cannot be reduced further, i.e., any further merging of cells would result in an invalid control cover.
We call such a cover \textit{maximally reduced}, see Definition \ref{def:maxreduced}. 
Reducedness of the control congruences obtained by Algorithm \ref{alg:localization} is addressed in Lemma \ref{lm:reducedness}.


\begin{definition}
\label{def:maxreduced}
Cover $\mathcal{C}_k$ is \textit{maximally reduced} w.r.t. $G,S,\Sigma_{c,k}$ iff
$\forall \pi_1,\pi_2\in \mathcal{C}_k \text{, if }\pi_1{\neq}\pi_2\text{, then }(\mathcal{C}_k\setminus \{\pi_1,\pi_2\}) \cup \{\pi_1\cup \pi_2\}$ is not a control congruence w.r.t. $G,S,\Sigma_{c,k}$.
\null\nobreak\hfill\ensuremath{\blacklozenge}
\end{definition}

\begin{lemma}
\label{lm:reducedness}
$\mathcal{C}_k$ obtained by Algorithm \ref{alg:localization}, is maximally reduced w.r.t. $G,S,\Sigma_{c,k}$. 

\begin{proof}
Algorithm \ref{alg:localization} iterates over all pairs of states, and only skips pairs of states when mergeability between some pair of states contained in the respective cells has already been checked.
Thus, if Algorithm \ref{alg:localization} outputs a control congruence containing individual cells $\pi_1$ and $\pi_2$, then mergeability has been checked between some pair of states $x_1\in \pi_1, x_2\in \pi_2$.
Let us say $x_1,x_2$ respectively were in cells $\rho_1,\rho_2$ at the point their mergeability was checked.
Since $x_1$ and $x_2$ were not merged, \texttt{check\_merge} has returned false for this evaluation, which means that some pair of states $x_3,x_4$ respectively in $\rho_1,\rho_2$ were not mergeable.
Since Algorithm \ref{alg:localization} only merges cells (i.e., never splits a cell), we know that for the resulting control congruence $x_3\in\rho_1\subseteq\pi_1$ and $x_4\in\rho_2\subseteq\pi_2$.
Since $x_3$ and $x_4$ are not mergeable, $\pi_1$ and $\pi_2$ cannot be merged to form a control congruence.
\end{proof}
\end{lemma}

Note that Lemma \ref{lm:reducedness} does not mean that the smallest control congruence is found by Algorithm \ref{alg:localization}.
A control congruence (and resulting local supervisor) is generally non-unique, and which is found by Algorithm \ref{alg:localization} depends on the order in which mergeability of the states is checked, which depends on their indexing.
Unfortunately, finding a control congruence with the smallest number of cells is an NP-hard problem \cite{Su2004}.

Lemmas \ref{lm:initialization} and \ref{lm:reducedness} are applicable for supervisor localization \cite{Cai2010} and supervisor reduction \cite{Su2004} (which also uses Algorithms \ref{alg:localization} and \ref{alg:checkmerge}, i.e., not only applicable to the transformational approach we present next.

%


\section{Problem definition}
\label{sec:problem}
We assume a base system $G$ consisting of $n$ agents, a supervisor $S$, and a partition $\dot\bigcup_{k\in \{1,...,n\}}\Sigma_{c,k}=\Sigma_c\subseteq \Sigma$ of controllable events.
This base system has been localized, i.e., a control congruence $\mathcal{C}_k$ was obtained for each agent $k$.

Now the system changes to variant system $G'$ consisting of $n'$ agents, a supervisor $S'$, and a partition of controllable events $\dot\bigcup_{k\in \{1,...,n'\}}\Sigma_{c,k}'=\Sigma_c'\subseteq \Sigma'$.
We compute $\mathcal{C}_k'$ and $LOC_k'$ for all $k$ from $1$ to $n'$ based on the control congruences of the base system, rather than starting localization from scratch. 
We call this procedure \textit{transformational supervisor localization} (TSL).
TSL is to correctly localize the variant system, as defined in Problem \ref{problem}.
Note that in this problem definition, any adaptation can be made to the base system (that generates a well-defined variant system).


\begin{problem}\label{problem}
Use $\mathcal{C}_k$ for $k$ from $1$ to $n$ of the base system $G,S$ to transformationally compute new local supervisors $LOC_k'$ for all $k$ from $1$ to $n'$ that are control equivalent (Equations \ref{eq:controlequiv1} and \ref{eq:controlequiv2}) to $S'$ with respect to $G'$.
\null\nobreak\hfill\ensuremath{\blacklozenge}
\end{problem}

Since a set of local supervisors can be constructed from a set of control covers, in our work we mainly focus on finding control covers (in this case, control congruences) for the variant system in a transformational approach.

%

Furthermore, it is desirable to have small local supervisors.
Therefore, TSL will compute maximally reduced control covers to use in the construction of the local supervisors.

\section{Transformational supervisor localization}
\label{sec:TSL}
In this section, we first discuss an algorithm that is used to transform a cover $\mathcal{C}_k$ to a control congruence in case the system has been adapted.
Next, we use this algorithm in the general procedure used for TSL.

\subsection{Isolating conflicts}
%

We consider the case that a control congruence $\mathcal{C}_k$ has been computed for some base system $S=(X,\Sigma,\xi,x_0,X_m)$, $G=(Q,\Sigma,\delta,q_0,Q_m)$.
Now the system is adapted to form variant system  $S'=(X',\Sigma',\xi',x_0',X_m')$, $G'=(Q',\Sigma',\delta',q_0',Q_m')$.
In our notation, we use $E'$, $D_k'$, ... to indicate that the function $E$, $D_k$, ... are applied to the variant automaton.
I.e., $E'$ is a function $E':X'\rightarrow 2^{\Sigma'}$.

\begin{algorithm}[t]
\small
 \caption{\texttt{isolate}}
 \label{alg:isolate}
 \begin{algorithmic}[1]
 \renewcommand{\algorithmicrequire}{\textbf{Input:}}
 \renewcommand{\algorithmicensure}{\textbf{Output:}}
\renewcommand{\algorithmicendif}{\textbf{end}}
\renewcommand{\algorithmicendfor}{\textbf{end}}
\renewcommand{\algorithmicendwhile}{\textbf{end}}
 \REQUIRE $\mathcal{C}_k$, $S$, $G'$, $S'$, $\Sigma_{c,k}'$
 \ENSURE $\mathcal{C}_k'$
 \STATE $\mathcal{C}_k'=\{\pi\setminus (X\setminus X')|\pi\in \mathcal{C}_k\}\cup\bigcup\{\{x\}|x\in X'\setminus X\}$
 \STATE $\mathit{flag}=\mathit{true}$
 \WHILE{$\mathit{flag}$}
 \STATE $\mathit{flag}=\mathit{false}$
 \FOR{\textbf{all} $x\in X'\cap X$}
 \IF{$\exists x'\in [x]_{\mathcal{C}_k'} : \big{(} (x,x')\not\in\mathcal{R}_k' \vee (\exists \sigma\in E'(x)\cap E'(x')) : ([\xi'(x,\sigma)]_{\mathcal{C}_k'} \neq [\xi'(x',\sigma)]_{\mathcal{C}_k'} )\big{)}$}
 \STATE $\mathit{flag}=\mathit{true}$
 \STATE $\mathcal{C}_k'=(\mathcal{C}_k'\setminus \{ [x]_{\mathcal{C}_k'} \})\cup \{[x]_{\mathcal{C}_k'}\setminus \{x\}\} \cup \{\{x\}\} $
 \ENDIF
 \ENDFOR
 \ENDWHILE
 \STATE \textbf{return} $\mathcal{C}'_k$
 \end{algorithmic}
\end{algorithm}

\begin{figure}[th]
\centering
\includegraphics[scale=1]{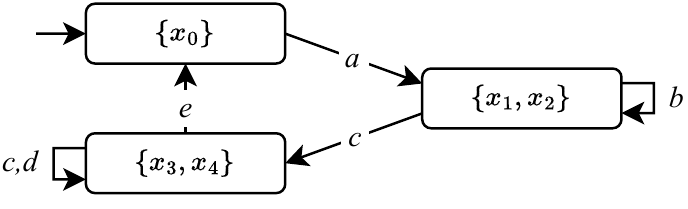}
\caption{Isolated state $x_0$}
\label{Ex2isolated}
\end{figure}

Algorithm \ref{alg:isolate} constructs a control congruence $\mathcal{C}_k'$ based on $\mathcal{C}_k$.
First, states that are removed from $X$ to create $X'$ are removed from the cells they were in in $\mathcal{C}_k$. 
New states are added as singleton cells.
Next, the algorithm looks for states $x$ that do not satisfy condition (\textit{i}) or (\textit{ii}) of a control cover from Section \ref{sec:preliminaries} anymore with a state $x'$ in the same cell.
If such a state $x$ is found, it is \textit{isolated}: it is removed from its initial cell and placed in a singleton cell. 
Note that conditions (\textit{i}) and (\textit{ii}) are always satisfied for states in a singleton cell.
Finally, all states that induce such a control consistency conflict are isolated, and the resulting cover is a control congruence.

We first present Example \ref{ex:ex2} to demonstrate the functioning of Algorithm \ref{alg:isolate}.
Next, we prove correctness of Algorithm \ref{alg:isolate} in Theorem \ref{th:isolate}.

\begin{example}
\label{ex:ex2}
Let us consider the case the system of Example 1 is adapted. 
In addition to the disablements $D_1(x_0){=}\{c\}, D_1(x_2){=}\{a\}$ in the base system, the variant system has an additional disablement: $D_1(x_3){=}\{a\}$.
As a result, for the variant system $(x_0,x_3)\not\in \mathcal{R}_1$.
Therefore, the cover found in Example \ref{ex:ex1} is not valid anymore. 
This conflict is found in line 6 of Algorithm \ref{alg:isolate}, and subsequently $x_0$ (or $x_3$ depending on order of iteration) is removed from its previous cell and placed in a singleton cell.
No more conflicts exist in the resulting cover.
Constructing a local supervisor for this cover yields the automaton shown in Fig. \ref{Ex2isolated}.
\null\nobreak\hfill\ensuremath{\blacklozenge}
\end{example}

\begin{theorem}
\label{th:isolate}
Given $N=|X\cap X'|$, Algorithm \ref{alg:isolate} terminates, has a worst-case time complexity of $\mathcal{O}(|\Sigma|{\cdot}N^3)$, and the generated cover $\mathcal{C}_k'$ is a control congruence w.r.t. $G',S',\Sigma_{c,k}'$.

\begin{proof}
A state in a singleton cell is trivially control consistent.
If in the for-loop (lines 5-10) a state is found that is not control consistent with another state in the same cell, it is placed in a singleton cell and removed from its original cell, and the algorithm iterates over all states in $X\cap X'$ again.
Eventually, since $N$ is finite, there are no more non-control consistent states, the for-loop terminates with $\mathit{flag=false}$, the while-loop breaks, and the algorithm terminates.

Checking the if-condition on line 6 has a worst-case cost of $|\Sigma|{\cdot}N$.
The for-loop (lines 5-10) is performed $N$ times in worst-case.
The while-loop (lines 3-11) is performed $N$ times in worst-case.
Therefore, the time complexity is $\mathcal{O}(|\Sigma|{\cdot}N^3)$.\footnote{To achieve this cost in implementation, instead of storing cells as state sets, a cell index number is stored for each state. A state can be isolated by simply assigning it with a new cell index. Since all cells are non-overlapping, comparing whether two cells are the same can be done by comparing the cell index of one state from each cell.}

The while-loop only breaks when conditions (\textit{i}) and (\textit{ii}) are both met for all states in $X\cap X'$.
Also all states in $X'\setminus X$ are control consistent as they are placed in singleton cells.
There is no overlap between cells in $\mathcal{C}_k'$ as all cells in $X'\setminus X$ are placed in singleton cells and no merges are performed for states in $X\cap X'$, which are initially partitioned by $\mathcal{C}_k$.
Thus, $\mathcal{C}_k'$ is a control congruence w.r.t. $G',S',\Sigma_{c,k}'$.
\end{proof}
\end{theorem}

\begin{algorithm}[t]
\small
 \caption{\texttt{TSL}}
 \label{alg:TSL}
 \begin{algorithmic}[1]
 \renewcommand{\algorithmicrequire}{\textbf{Input:}}
 \renewcommand{\algorithmicensure}{\textbf{Output:}}
\renewcommand{\algorithmicendif}{\textbf{end}}
\renewcommand{\algorithmicendfor}{\textbf{end}}
\renewcommand{\algorithmicendwhile}{\textbf{end}}
 \REQUIRE $\{\mathcal{C}_k|1\leq k\leq n\}$, $S$, $G'$, $S'$, $\{\Sigma_{c,k}'|1\leq k\leq n'\}$, $\mathcal{M}$
 \ENSURE $\{LOC_k'|1\leq k\leq n'\}$, $\{C_k'|1\leq k\leq n'\}$
 \FOR{$k=1$ to $n'$}
 \IF{$\mathcal{M}(k)\neq 0$}
 \STATE $\mathcal{C}_{k,init}'=\text{\texttt{isolate}}(\mathcal{C}_{\mathcal{M}(k)},S,G',S',\Sigma_{c,k}')$
 \ELSE
 \STATE $\mathcal{C}_{k,init}'=\{\{x\}|x\in X'\}$
 \ENDIF
 \STATE $\mathcal{C}_k'=\text{\texttt{localize}}(G',S',\Sigma_{c,k}',\mathcal{C}_{k,init}')$ 
 \STATE Compute $LOC_k'$ based on $\mathcal{C}_k'$
 \ENDFOR
 \STATE \textbf{return} $\{LOC_k'|1\leq k\leq n'\}$, $\{C_k'|1\leq k\leq n'\}$
 \end{algorithmic}
\end{algorithm}

\subsection{General procedure}
\label{sec:generalprocedure}
In this section we present the TSL procedure, show in Theorem \ref{th:TSL} that TSL solves Problem \ref{problem}, and in Theorem \ref{th:reducedness} that the resulting control congruences are maximally reduced.

The TSL procedure is sketched in pseudo-code in Algorithm \ref{alg:TSL}.
We assume a mapping $\mathcal{M}:\{1,...,n'\} \rightarrow \{0,...,n\}$, that maps every agent in the variant system to either an agent of the base system, or to `$0$'.
If $\mathcal{M}(k){=}0$, it means no base control cover is selected and the inital control congruence is set to a singleton cover.
In case $\mathcal{M}(k)$ is nonzero, control congruence $\mathcal{C}_{\mathcal{M}(k)}$ is selected from the base system to perform \texttt{isolate} to find an initial control congruence.
After performing \texttt{isolate}, the resulting cover might not be maximally reduced.
This is why, after performing \texttt{isolate}, the cover is used to initialize \texttt{localize} in order to merge cells whenever possible.
The reasoning for the TSL procedure is that \texttt{isolate} produces a control congruence in which generally states will already be merged into cells, limiting the work that needs to be done during \texttt{localize}.
This is demonstrated in Example \ref{ex:ex3}.
\texttt{TSL} also returns covers $\{C_k'|1\leq k\leq n'\}$ so that they can be used in a next TSL if the system is further adapted.

\begin{figure}[t]
\centering
\includegraphics[scale=1]{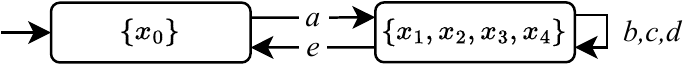}
\caption{Local supervisor of variant system}
\label{Ex2LOC}
\end{figure}

\begin{example}
\label{ex:ex3}
This is a continuation of Example \ref{ex:ex2}, in which a variant system was presented to the base system of Example \ref{ex:ex1}, and \texttt{isolate} was performed to compute a control congruence for the variant system, yielding the local supervisor of Fig. \ref{Ex2isolated}.
However, the cover can be further reduced, resulting in a local supervisor with less states.
We perform \texttt{localize} initialized with the cover found in Example \ref{ex:ex2}.
$\{x_0\}$ cannot merge with $\{x_1,x_2\}$ for multiple reasons: $x_1$ and $x_2$ both enable event $c$, which is disabled in $x_0$, and $x_0$ enables event $a$, which is disabled in $x_2$.
$\{x_0\}$ cannot merge with $\{x_3,x_4\}$ as $x_0$ enables $a$, which is disabled in $x_3$ in the variant system.
$\{x_1,x_2\}$ can be merged with $\{x_3,x_4\}$: there are no conflicts.
After merging these cells, no further merges are possible, leading to control congruence $\{\{x_0\},\{x_1,x_2,x_3,x_4\}\}$. 
Constructing a local supervisor for this cover yields the automaton in Fig. \ref{Ex2LOC}.
\null\nobreak\hfill\ensuremath{\blacklozenge}
\end{example}

\begin{theorem}
\label{th:TSL}
Algorithm \ref{alg:TSL} terminates, has worst-case complexity $\mathcal{O}(n'{\cdot}|\Sigma'|{\cdot}|X'|^4)$, and solves Problem \ref{problem}.

\begin{proof}
Algorithm \ref{alg:TSL} terminates because \texttt{isolate} (Theorem \ref{th:isolate}) and \texttt{localize} (\cite{Su2004,Cai2010}) terminate.

In worst-case, \texttt{isolate} is called $n'$ times, and its complexity is $\mathcal{O}(|\Sigma'|{\cdot}|X\cap X'|^3)$ (Theorem \ref{th:isolate}).
\texttt{localize} is called $n'$ times, and its complexity is $\mathcal{O}(|\Sigma'|{\cdot}|X'|^4)$ \cite{Su2004,Cai2010}. 
Therefore, the complexity of \texttt{TSL} is $\mathcal{O}(n'{\cdot}|\Sigma'|{\cdot}|X'|^4)$.

For each agent in the variant system, \texttt{localize} is initiated with a control congruence, since line 3 constructs a control congruence (Theorem \ref{th:isolate}) and the singleton cover constructed in line 5 is a control congruence.
Thus, the covers computed by \texttt{localize} are control congruences following from Lemma \ref{lm:initialization}.
It is shown in \cite{Cai2010} that local supervisors constructed from control congruences satisfy Problem \ref{problem}.
\end{proof}
\end{theorem}

Clearly, SL and TSL have the same complexity. 
The idea is that TSL is quicker in practice, when the variant system is sufficiently similar to the base system.
Unfortunately, at the moment we can not predict whether TSL will be quicker than SL.
We present some experiments in Section \ref{sec:CMT} to study the computational benefit in practice.

In addition to correctness of the result, \texttt{TSL} also produces maximally reduced control congruences.
\begin{theorem}
\label{th:reducedness}
All $\mathcal{C}_k'\in \{\mathcal{C}_k'|1\leq k\leq n'\}$ obtained by Algorithm \ref{alg:TSL} are maximally reduced w.r.t. $G',S',\Sigma_{c,k}'$. 

\begin{proof}
Every control congruence $\mathcal{C}_k'$ that is returned by Algorithm \ref{alg:TSL} is constructed by performing Algorithm \ref{alg:localization}.
Control congruences constructed by Algorithm \ref{alg:localization} are maximally reduced (Lemma \ref{lm:reducedness}).
Thus, the theorem holds.
\end{proof}
\end{theorem}

\section{Case study: Cat and Mouse Tower}
\label{sec:CMT}
\begin{figure}[t]
      \centering
        \includegraphics[scale=0.8]{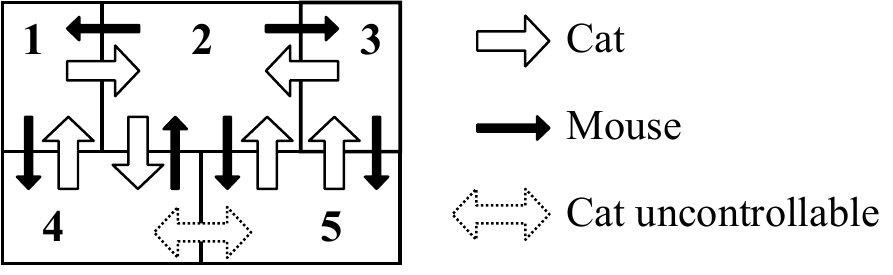}
        \caption{CMT room layout of a level \cite{Thuijsman2021}}
        \label{fig:CMT}
    \end{figure}

As a case study to evaluate the efficiency of TSL relative to SL, we take the Cat and Mouse Tower (CMT) from \cite{Ma2008}.
There are $n$ floors, and on each floor of the tower there are five rooms as shown in Fig. \ref{fig:CMT}.
Cats and mice can move between the rooms as indicated by the arrows.
Between each level there is a connection for both cats and mice.
This connection is between room $j$ of level $5\cdot i+j$ to room $j$ of level $5\cdot i+j+1$, for $i\in \mathbb{N}_0$, $j\in \{1,2,3,4,5\}$, and $5\cdot i+j < n$.
So room 1 level 1 is connected to room 1 level 2; room 2 level 2 is connected to room 2 level 3; and so forth, essentially forming a spiraling staircase.
All doors can be controlled, except for the bidirectional cat door between rooms 2 and 4.
There are $k$ cats and $k$ mice, and consequently each room can also hold between $0$ and $k$ cats and/or mice.
The cats start in room $1$ of level $1$, and the mice start in room $5$ of level $n$.
The requirement of this system is that there can never be a cat and a mouse in the same room at the same time.

\begin{table*}[t]
\caption{CMT experimental results}
\label{tab:CMT}
\begin{center}
\small
\begin{tabular}{ll||rrrrr||rrrr|}
\cline{3-11}
                                                                                                                &                                                      & \multicolumn{5}{c|}{mean runtime}                                                                                                                                                                                                                                      & \multicolumn{4}{c|}{mean \# cells}                                                                                                              \\ \hline
\multicolumn{1}{|l|}{variant system}                                                                            & agent & \multicolumn{1}{r|}{SL [s]} & \multicolumn{1}{r|}{\texttt{isolate} [s]} & \multicolumn{1}{r|}{\begin{tabular}[c]{@{}r@{}}initialized \\ \texttt{localize} [s]\end{tabular}} & \multicolumn{1}{r|}{\begin{tabular}[c]{@{}r@{}}TSL [s]\\ (sum)\end{tabular}} & \% change & \multicolumn{1}{r|}{SL}   & \multicolumn{1}{r|}{\begin{tabular}[c]{@{}r@{}}initial\\ guess\end{tabular}} & \multicolumn{1}{r|}{isolated} & TSL  \\ \specialrule{1.1pt}{0pt}{0pt}
\multicolumn{1}{|l|}{\multirow{4}{*}{\begin{tabular}[c]{@{}l@{}}1\\ (362 states,\\ 1142 trans.)\end{tabular}}}  & 1                                                    & \multicolumn{1}{r|}{1.22}   & \multicolumn{1}{r|}{0.05}                 & \multicolumn{1}{r|}{0.15}                                                                         & \multicolumn{1}{r|}{0.20}                                                    & -83\%     & \multicolumn{1}{r|}{11.6} & \multicolumn{1}{r|}{11.6}                                                    & \multicolumn{1}{r|}{11.6}     & 11.6 \\ \cline{2-11} 
\multicolumn{1}{|l|}{}                                                                                          & 2                                                    & \multicolumn{1}{r|}{1.11}   & \multicolumn{1}{r|}{0.04}                 & \multicolumn{1}{r|}{0.21}                                                                         & \multicolumn{1}{r|}{0.25}                                                    & -78\%     & \multicolumn{1}{r|}{14.3} & \multicolumn{1}{r|}{14.3}                                                    & \multicolumn{1}{r|}{14.3}     & 14.3 \\ \cline{2-11} 
\multicolumn{1}{|l|}{}                                                                                          & 3                                                    & \multicolumn{1}{r|}{1.16}   & \multicolumn{1}{r|}{0.04}                 & \multicolumn{1}{r|}{0.15}                                                                         & \multicolumn{1}{r|}{0.18}                                                    & -84\%     & \multicolumn{1}{r|}{13.7} & \multicolumn{1}{r|}{13.7}                                                    & \multicolumn{1}{r|}{13.7}     & 13.7 \\ \cline{2-11} 
\multicolumn{1}{|l|}{}                                                                                          & 4                                                    & \multicolumn{1}{r|}{2.47}   & \multicolumn{1}{r|}{0.06}                 & \multicolumn{1}{r|}{0.08}                                                                         & \multicolumn{1}{r|}{0.14}                                                    & -94\%     & \multicolumn{1}{r|}{10.9} & \multicolumn{1}{r|}{10.9}                                                    & \multicolumn{1}{r|}{10.9}     & 10.9 \\ \specialrule{1.1pt}{0pt}{0pt}
\multicolumn{1}{|l|}{\multirow{4}{*}{\begin{tabular}[c]{@{}l@{}}2 \\ (375 states,\\ 1214 trans.)\end{tabular}}} & 1                                                    & \multicolumn{1}{r|}{1.84}   & \multicolumn{1}{r|}{0.03}                 & \multicolumn{1}{r|}{1.73}                                                                         & \multicolumn{1}{r|}{1.76}                                                    & -4\%      & \multicolumn{1}{r|}{11.5} & \multicolumn{1}{r|}{23.9}                                                    & \multicolumn{1}{r|}{295.6}    & 18.8 \\ \cline{2-11} 
\multicolumn{1}{|l|}{}                                                                                          & 2                                                    & \multicolumn{1}{r|}{1.10}   & \multicolumn{1}{r|}{0.05}                 & \multicolumn{1}{r|}{1.26}                                                                         & \multicolumn{1}{r|}{1.30}                                                    & +18\%      & \multicolumn{1}{r|}{15.0} & \multicolumn{1}{r|}{27.8}                                                    & \multicolumn{1}{r|}{231.6}    & 22.1 \\ \cline{2-11} 
\multicolumn{1}{|l|}{}                                                                                          & 3                                                    & \multicolumn{1}{r|}{1.32}   & \multicolumn{1}{r|}{0.04}                 & \multicolumn{1}{r|}{1.21}                                                                         & \multicolumn{1}{r|}{1.26}                                                    & -5\%      & \multicolumn{1}{r|}{14.7} & \multicolumn{1}{r|}{27.5}                                                    & \multicolumn{1}{r|}{232.7}    & 21.7 \\ \cline{2-11} 
\multicolumn{1}{|l|}{}                                                                                          & 4                                                    & \multicolumn{1}{r|}{3.78}   & \multicolumn{1}{r|}{0.03}                 & \multicolumn{1}{r|}{1.92}                                                                         & \multicolumn{1}{r|}{1.95}                                                    & -48\%     & \multicolumn{1}{r|}{11.1} & \multicolumn{1}{r|}{24.3}                                                    & \multicolumn{1}{r|}{292.7}    & 17.2 \\ \specialrule{1.1pt}{0pt}{0pt}
\multicolumn{1}{|l|}{\multirow{4}{*}{\begin{tabular}[c]{@{}l@{}}3\\ (270 states,\\ 853 trans.)\end{tabular}}}   & 1                                                    & \multicolumn{1}{r|}{0.78}   & \multicolumn{1}{r|}{0.04}                 & \multicolumn{1}{r|}{0.10}                                                                         & \multicolumn{1}{r|}{0.14}                                                    & -82\%     & \multicolumn{1}{r|}{9.4}  & \multicolumn{1}{r|}{8.8}                                                     & \multicolumn{1}{r|}{8.8}      & 8.8  \\ \cline{2-11} 
\multicolumn{1}{|l|}{}                                                                                          & 2                                                    & \multicolumn{1}{r|}{0.71}   & \multicolumn{1}{r|}{0.02}                 & \multicolumn{1}{r|}{0.09}                                                                         & \multicolumn{1}{r|}{0.11}                                                    & -85\%     & \multicolumn{1}{r|}{12.8} & \multicolumn{1}{r|}{14.1}                                                    & \multicolumn{1}{r|}{14.1}     & 13.9 \\ \cline{2-11} 
\multicolumn{1}{|l|}{}                                                                                          & 3                                                    & \multicolumn{1}{r|}{0.88}   & \multicolumn{1}{r|}{0.03}                 & \multicolumn{1}{r|}{0.10}                                                                         & \multicolumn{1}{r|}{0.12}                                                    & -86\%     & \multicolumn{1}{r|}{10.8} & \multicolumn{1}{r|}{14.3}                                                    & \multicolumn{1}{r|}{14.7}     & 13.9 \\ \cline{2-11} 
\multicolumn{1}{|l|}{}                                                                                          & 4                                                    & \multicolumn{1}{r|}{6.90}   & \multicolumn{1}{r|}{0.04}                 & \multicolumn{1}{r|}{3.68}                                                                         & \multicolumn{1}{r|}{3.72}                                                    & -46\%     & \multicolumn{1}{r|}{2.5}  & \multicolumn{1}{r|}{10.0}                                                    & \multicolumn{1}{r|}{10.0}     & 9.0  \\ \specialrule{1.1pt}{0pt}{0pt}
\multicolumn{1}{|l|}{\multirow{4}{*}{\begin{tabular}[c]{@{}l@{}}4\\ (309 states,\\ 986 trans.)\end{tabular}}}   & 1                                                    & \multicolumn{1}{r|}{2.19}   & \multicolumn{1}{r|}{0.05}                 & \multicolumn{1}{r|}{0.65}                                                                         & \multicolumn{1}{r|}{0.69}                                                    & -68\%     & \multicolumn{1}{r|}{8.9}  & \multicolumn{1}{r|}{11.2}                                                    & \multicolumn{1}{r|}{11.2}     & 10.6 \\ \cline{2-11} 
\multicolumn{1}{|l|}{}                                                                                          & 2                                                    & \multicolumn{1}{r|}{0.82}   & \multicolumn{1}{r|}{0.03}                 & \multicolumn{1}{r|}{0.10}                                                                         & \multicolumn{1}{r|}{0.12}                                                    & -85\%     & \multicolumn{1}{r|}{12.8} & \multicolumn{1}{r|}{14.3}                                                    & \multicolumn{1}{r|}{14.3}     & 14.3 \\ \cline{2-11} 
\multicolumn{1}{|l|}{}                                                                                          & 3                                                    & \multicolumn{1}{r|}{0.79}   & \multicolumn{1}{r|}{0.03}                 & \multicolumn{1}{r|}{0.15}                                                                         & \multicolumn{1}{r|}{0.17}                                                    & -78\%     & \multicolumn{1}{r|}{13.9} & \multicolumn{1}{r|}{14.1}                                                    & \multicolumn{1}{r|}{14.1}     & 14.1 \\ \cline{2-11} 
\multicolumn{1}{|l|}{}                                                                                          & 4                                                    & \multicolumn{1}{r|}{1.78}   & \multicolumn{1}{r|}{0.04}                 & \multicolumn{1}{r|}{0.08}                                                                         & \multicolumn{1}{r|}{0.13}                                                    & -93\%     & \multicolumn{1}{r|}{10.6} & \multicolumn{1}{r|}{11.4}                                                    & \multicolumn{1}{r|}{11.4}     & 11.4 \\ \specialrule{1.1pt}{0pt}{0pt}
\multicolumn{1}{|l|}{\multirow{4}{*}{\begin{tabular}[c]{@{}l@{}}5\\ (403 states,\\ 1304 trans.)\end{tabular}}}  & 1                                                    & \multicolumn{1}{r|}{2.23}   & \multicolumn{1}{r|}{0.05}                 & \multicolumn{1}{r|}{1.49}                                                                         & \multicolumn{1}{r|}{1.54}                                                    & -31\%     & \multicolumn{1}{r|}{12.1} & \multicolumn{1}{r|}{51.5}                                                    & \multicolumn{1}{r|}{327.4}    & 14.7 \\ \cline{2-11} 
\multicolumn{1}{|l|}{}                                                                                          & 2                                                    & \multicolumn{1}{r|}{1.41}   & \multicolumn{1}{r|}{0.05}                 & \multicolumn{1}{r|}{0.66}                                                                         & \multicolumn{1}{r|}{0.71}                                                    & -50\%     & \multicolumn{1}{r|}{14.6} & \multicolumn{1}{r|}{55.2}                                                    & \multicolumn{1}{r|}{269.4}    & 15.9 \\ \cline{2-11} 
\multicolumn{1}{|l|}{}                                                                                          & 3                                                    & \multicolumn{1}{r|}{1.79}   & \multicolumn{1}{r|}{0.05}                 & \multicolumn{1}{r|}{0.61}                                                                         & \multicolumn{1}{r|}{0.66}                                                    & -63\%     & \multicolumn{1}{r|}{14.3} & \multicolumn{1}{r|}{55.1}                                                    & \multicolumn{1}{r|}{262.0}    & 14.4 \\ \cline{2-11} 
\multicolumn{1}{|l|}{}                                                                                          & 4                                                    & \multicolumn{1}{r|}{4.59}   & \multicolumn{1}{r|}{0.03}                 & \multicolumn{1}{r|}{1.49}                                                                         & \multicolumn{1}{r|}{1.52}                                                    & -67\%     & \multicolumn{1}{r|}{11.2} & \multicolumn{1}{r|}{52.4}                                                    & \multicolumn{1}{r|}{323.2}    & 11.7 \\ \hline
\end{tabular}
\end{center}
\end{table*}

As base system, we take a tower with four levels, one cat, and one mouse.
The monolithic supervisor of this system has 362 states and 1159 transitions.
For localization, we consider each level as a separate agent. 
An agent controls all events of the cat and mouse that originate in that level, e.g., the level 1 agent controls all doors on that level, and the movements from level 1 room 1 to level 2 room 1 (but not the other way around; these are controlled by the level 2 agent). 

We construct five variant systems (each modifies the base system directly, i.e., the adaptations are not cumulative):
\begin{enumerate}
\item Removed cat door from room 3 to room 4 on level 2. 
\item Made all doors controllable.
\item Added requirement that cats should never reach level 4. 
\item Removed room 5 of level 1. 
\item Added a room 6 to level 1 with bidirectional controllable doors for cat and mouse to room 5 of level 1.
\end{enumerate}

The models and a proof-of-concept implementation of the algorithms have been made in Matlab\footnote{All used models and algorithms can be found here: {\url{https://github.com/sbthuijsman/TSL}}.}. 
We performed SL for the base system, and SL and TSL for each variant system.
For TSL, each agent (floor) of the variant system is mapped to the same floor in the base system.
A standard personal computer with i7 processor was used.
Matlab used less than 2 GB of memory.
Since we draw conclusions on relative and not absolute runtimes, the conclusions are not influenced by the hardware.
Because the results are influenced by state indexing order, the experiments are performed for ten random index orders and mean values over those runs are presented. 

In the left side of Table \ref{tab:CMT} we compare the computation time in seconds of performing SL and TSL for the agents in the variant system.
To provide further detail, we show how much time of performing TSL is spent on the \texttt{isolate} and \texttt{localize} portion of the procedure.
The percentage change comparing TSL to SL is displayed, where a negative or positive value respectively indicates how much quicker or slower TSL is compared to SL.

In the right side of Table \ref{tab:CMT} we compare the number of cells between the result of SL and TSL for the agents in the variant system.
The numbers under `initial guess' indicate the number of cells of $\mathcal{C}_k'$ after line 1 of \texttt{isolate}, before any states are isolated.
The numbers under `isolated' indicate the number of cells after completing \texttt{isolate}, but before \texttt{localize} is performed.

For the first variant system, we observe that no states need to be isolated during \texttt{isolate} and no further merges of cells can be performed when performing \texttt{localize} initialized by the control cover of the base system.
Compared to performing SL initialized by a singleton cover, TSL is much quicker.
For the second variant system, there is much less computational benefit.
Here, a local system is found were TSL is slower than SL, i.e., in this case localization is quicker when initialized by a singleton cover.
At the moment, we have no way to predict when this will be the case.
We observe that for this system a lot of states need to be isolated for all subsystems.
Even so, isolation is performed relatively quickly.
Because the isolated cover is relatively close to the singleton cover (which has 375 cells), TSL runtimes are relatively close to the SL runtimes.
Another observation is that TSL computes covers with more cells than SL, because it starts with a coarser cover which limits the cell merges that can be made during \texttt{localize}.
For variant systems 3, 4, and 5 TSL is consistently quicker than SL, even though for variant system 5 a lot of states require to be isolated.

The same experiments have been performed for larger instances of CMT, with 6 levels (842 states) and 8 levels (1525 states). 
Because of space constraints we cannot fully present those results in this paper, they are available in the repository linked above.
The same conclusions can be made for these larger instances.
Respectively, the average percentage change over all local systems for CMT with 4, 6, and 8 levels, were $-61\%$, $-54\%$, and $-57\%$.

From a monolithic point of view the adaptations made to the CMT system are considerable (reflected in the change in number of states and transitions).
Regardless, these experiments suggest that TSL is more efficient than performing SL from scratch.

\section{Conclusions}
\label{sec:conclusion}
We presented a TSL procedure, that reuses control congruences from a previous SL to more efficiently compute these control congruences for a system once it is adapted.
Correctness of the algorithms is shown, and examples are provided.
The method is evaluated by means of some experiments on the CMT system.
For these experiments, the runtime of TSL is shown to be lower than SL.

\bibliography{references}

\begin{thebibliography}{1}
\providecommand{\url}[1]{#1}
\csname url@rmstyle\endcsname
\providecommand{\newblock}{\relax}
\providecommand{\bibinfo}[2]{#2}
\providecommand\BIBentrySTDinterwordspacing{\spaceskip=0pt\relax}
\providecommand\BIBentryALTinterwordstretchfactor{4}
\providecommand\BIBentryALTinterwordspacing{\spaceskip=\fontdimen2\font plus
\BIBentryALTinterwordstretchfactor\fontdimen3\font minus
  \fontdimen4\font\relax}
\providecommand\BIBforeignlanguage[2]{{%
\expandafter\ifx\csname l@#1\endcsname\relax
\typeout{** WARNING: IEEEtran.bst: No hyphenation pattern has been}%
\typeout{** loaded for the language `#1'. Using the pattern for}%
\typeout{** the default language instead.}%
\else
\language=\csname l@#1\endcsname
\fi
#2}}

\bibitem{Ramadge1987}
P.~Ramadge and W.~Wonham, ``Supervisory control of a class of discrete event
  processes,'' \emph{SIAM J. Control}, vol.~25, no.~1, pp. 206--230, 1987.

\bibitem{Cassandras2021}
C.~Cassandras and S.~Lafortune, \emph{Introduction to Discrete Event Systems},
  3rd~ed.\hskip 1em plus 0.5em minus 0.4em\relax Springer, 2021.

\bibitem{Moormann2021}
L.~Moormann, R.~Schouten, J.~{van de Mortel-Fronczak}, W.~Fokkink, and
  J.~Rooda, ``Synthesis and implementation of distributed supervisory
  controllers with communication delays,'' in \emph{Conf. Autom. Sci.
  Eng.}\hskip 1em plus 0.5em minus 0.4em\relax {IEEE}, 2021.

\bibitem{Cai2010}
K.~Cai and W.~Wonham, ``Supervisor localization: A top-down approach to
  distributed control of discrete-event systems,'' \emph{Trans. Autom.
  Control}, vol.~55, no.~3, pp. 605--618, 2010.

\bibitem{Su2004}
R.~Su and W.~Wonham, ``Supervisor reduction for discrete-event systems,''
  \emph{Discrete Event Dynamic Syst.}, vol.~14, no.~1, pp. 31--53, 2004.

\bibitem{Thuijsman2022}
S.~Thuijsman and M.~Reniers, ``Transformational supervisor synthesis for
  evolving systems,'' \emph{Discrete Event Dynamic Syst.}, vol.~32, no.~2, pp.
  317--358, 2022.

\bibitem{Thuijsman2022b}
S.~Thuijsman, M.~Reniers, and K.~Cai, ``Transformational nonblocking
  verification,'' \emph{{IFAC}-{PapersOnLine}}, vol.~55, no.~28, pp. 256--263,
  2022.

\bibitem{Ma2008}
C.~{Ma} and W.~{Wonham}, ``{STSLib} and its application to two benchmarks,'' in
  \emph{IEEE Workshop Discrete Event Syst.}, 2008, pp. 119--124.

\bibitem{Thuijsman2021}
S.~Thuijsman, M.~Reniers, and D.~Hendriks, ``Efficiently enforcing mutual state
  exclusion requirements in symbolic supervisor synthesis,'' in \emph{Conf.
  Autom. Sci. Eng.}\hskip 1em plus 0.5em minus 0.4em\relax {IEEE}, 2021.

\end{thebibliography}

\end{document}